\documentclass[twocolumn,aps,prl,superscriptaddress,longbibliography]{revtex4-2}
\usepackage{graphicx}
\usepackage{color}
\usepackage{amsmath,amsfonts,amssymb,bm}
\usepackage[svgnames]{xcolor}
\usepackage [latin1]{inputenc}
\usepackage[colorlinks,linkcolor=MediumBlue,citecolor=MediumBlue,urlcolor = MediumBlue]{hyperref}
\usepackage{url}
\usepackage{stackengine}
\usepackage{eufrak}

\newcommand\barbelow[1]{\stackunder[1.2pt]{$#1$}{\rule{.8ex}{.075ex}}}

\graphicspath{{./}}

\newcommand{\sps}{\textsl{\textsc{s}}}
\newcommand{\su}{\textsl{\textsc{u}}}
\newcommand{\sv}{\textsl{\textsc{v}}}
\newcommand{\sw}{\textsl{\textsc{w}}}
\newcommand{\sa}{\textsl{\textsc{a}}}
\newcommand{\spb}{\textsl{\textsc{b}}}
\newcommand{\spt}{\textsl{\textsc{t}}}

\begin{document}

\title{Breakdown of Emergent Chiral Order and Defect Chaos in Nonreciprocal Flocks}

\author{Charlotte Myin}%
\email{charlotte.myin@ds.mpg.de}
\affiliation{Max Planck Institute for Dynamics and Self-Organization (MPI-DS), 37077 G\"ottingen, Germany}

\author{Suropriya Saha}%
\email{suropriya.saha@ds.mpg.de}
\affiliation{Max Planck Institute for Dynamics and Self-Organization (MPI-DS), 37077 G\"ottingen, Germany}

\author{Beno\^it Mahault}%
\email{benoit.mahault@umontpellier.fr}
\affiliation{Laboratoire Charles Coulomb (L2C), UMR 5221 CNRS--Universit{\'e} de Montpellier, Montpellier F-34095, France}
\affiliation{Max Planck Institute for Dynamics and Self-Organization (MPI-DS), 37077 G\"ottingen, Germany}
\date{\today}%

\begin{abstract}
We show that chiral order in two-dimensional nonreciprocal flocking mixtures is generically unstable. Combining large-scale agent-based simulations with a coarse-grained continuum description, we demonstrate that rotating chiral states emerging from antisymmetric couplings are destroyed by the proliferation of topological defects. The resulting dynamics is spatiotemporally chaotic and characterized by a finite correlation length that diverges as nonreciprocity vanishes. On length scales below this cutoff, density and orientational order fluctuations remain scale-free, but the associated scaling exhibits nonuniversal exponents. We attribute this atypical behavior to the coupling between density and order, which causes topological defects to act as persistent sources of nonlinear fluctuations.
\end{abstract}

\maketitle

%%%%%%%%%%%%%%%%%%%%%%%%%%%%%%%%%%%
%           Introduction
%%%%%%%%%%%%%%%%%%%%%%%%%%%%%%%%%%%

A central question in condensed matter physics is whether emergent order can persist at macroscopic scales.
At equilibrium, the breaking of a continuous symmetry produces massless Goldstone modes whose fluctuations, as formalized by the Hohenberg-Mermin-Wagner theorem, destroy long-range order in dimensions $d \le 2$~\cite{mermin1966,Hohenberg1967}.
Systems driven far from equilibrium such as active matter can circumvent this constraint.
Assemblies of coherently moving agents, or flocks, exhibit true long-range polar order in $d \ge 2$~\cite{Vicsek,TonerPRL95,tonerPRE98,tonerPRE2012,mahaultPRL2019,chatePRL2024,jentschPRL2024,chen2025} and are a paradigmatic, but not unique~\cite{ginelliPRL2010,mahaultPRL2021,dadhichiPRE2020,loosPRL2023,dopieralaPRL2025,pisegnaPNAS2024,ketaSM2025}, example.
Although robust to smooth, long-wavelength fluctuations, polar order in flocks has recently been shown to be particularly vulnerable to rare but large fluctuations inducing nucleation of counter-propagating fronts.
The flocking phase can then become partially~\cite{codinaPRL2022} or entirely~\cite{bessePRL2022,benvegnenPRL2023,wooPRL2024,dopieralaPRL2025,popliPRL2025} metastable.

Nonreciprocal mixtures---which can be described in terms of effective interactions breaking action-reaction symmetry---constitute another particularly rich class of active matter~\cite{BowickPRX2022,fruchart2026}.
One of their hallmarks is the emergence of oscillatory instabilities arising through the crossing of an exceptional point~\cite{saha2020scalar,you2020nonreciprocity,Fruchart2021Apr}.
For systems subject to conservation laws, these instabilities give rise to a variety of dynamical patterns~\cite{agudo2019active,saha2020scalar,you2020nonreciprocity,frohoffPRE2021,duan2023dynamical,dinelli2023non,braunsPRX2024,ranaPRL2024,duanPRR2025,grevePRL2025,sahaNC2025},
whereas nonconserved dynamics displays global oscillations and persistent rotations~\cite{Fruchart2021Apr,hanaiPRX2024,KreienkampPRL2024,KreienkampPRE2024,guislainPRE2024,guislainJSTAT2024,chen2024emergent,kreienkampComPhys2025,avni2025nonreciprocal,avni2025dynamical,martin2025transition,blomPRE2025}.

In multispecies flocks with nonreciprocal interactions, global rotations of the order parameter are associated with the spontaneous breaking of both chiral and time-translation symmetries. 
General arguments suggest that the associated Goldstone mode obeys a compact Kardar-Parisi-Zhang (KPZ) equation~\cite{kardarPRL1986,grinsteinPRL1993,chatePRL1995,pisegnaPNAS2024,maitraAnnRev2025,DavietPRL2025}. 
A major consequence is that, since KPZ interfaces are rough in all dimensions $d\le2$~\cite{kardarPRL1986,canetPRL2010}, fluctuations of the Goldstone mode suppress true long-range chiral order. 
More dramatically, in $d\le2$ compact KPZ dynamics allows for the proliferation of topological defects that preclude any long-range correlations~\cite{wachtelPRB2016,avni2025nonreciprocal,avni2025dynamical},
while scale-free behavior should persist only below a cutoff scale set by the mean inter-defect separation~\cite{deligiannisPRR2022,DavietPRL2025}. 
These arguments, however, do not account for the fact that nonreciprocal flocks are also characterized by several conserved density fields~\cite{Fruchart2021Apr,myin2026,DADAM,Maitra_PhysRevLett.125.238005,maitraAnnRev2025}. 
How the inherent coupling between density and polarity 
affects the stability of order and scaling properties in such active nonreciprocal mixtures remains so far unexplored.

In this Letter, we investigate the robustness of spontaneously emerging global chiral order in nonreciprocal flocks.
Combining large-scale agent-based simulations with a coarse-grained continuum theory,
our results confirm that chiral order and long-range correlations are generically destroyed by the proliferation of topological defects.
The resulting highly dynamical defect-chaos phase is dominated by strong spatiotemporal fluctuations,
where both orientational and density correlations exhibit scale-free behavior over intermediate length scales.
Remarkably, the associated scaling exponents are nonuniversal and incompatible with those of the KPZ universality class.
We attribute these anomalous scaling properties to the coupling between density and orientational order,
by which topological defects act as nucleation sites for highly nonlinear perturbations that constantly remodel both density and polarity landscapes.

%%%%%%%%%%%%%%%%%%%%%%%%%%%%%%%%%%%
%      Model and phase diagram
%%%%%%%%%%%%%%%%%%%%%%%%%%%%%%%%%%%

\textit{Minimal model for nonreciprocal flocks---}
We consider a binary mixture of Vicsek-like particles~\cite{Vicsek,myin2026} moving at constant speed on a two-dimensional plane. Particles align their velocities with neighbors located within a unit-radius disk, in the presence of angular noise.
The two species are labeled $\sa$ and $\spb$. A particle $k$ of species $\sps$ is characterized at discrete time $t$ by its position $\bm r_k^{\sps}(t)$ and its self-propulsion orientation $\theta_k^{\sps}(t)$, which evolve according to
\begin{subequations}
	\begin{align}
		\label{micromodel_r}
		\bm r_k^{\sps}(t + 1) & = \bm r_k^{\sps}(t) + v_0 \hat{\bm u}[\theta_k^{\sps}(t+1)],\\
		\label{micromodel_t}
		\theta_k^{\sps}(t + 1) & = {\rm arg}\bigg[ \sum_{\su, j} \chi^{\sps \su} e^{i \theta_j^{\su}(t)} 
		+ \eta n_k(t) e^{i \xi_k(t)} \bigg] ,
	\end{align}
	\label{micromodel}
\end{subequations}
where $\hat{\bm u}[\theta] = (\cos\theta \, \sin\theta)^T$, 
$v_0$ is the self-propulsion speed and $\eta$ the noise strength, which are taken identical for both species. 
The sum in Eq.~\eqref{micromodel_t} runs over all $n_k(t)$ neighbors $j$ of species $\su = \sa,\spb$.

The aligning interactions are encoded in the matrix $\chi^{\sps\su}$. 
For $\chi^{\sps\su} > 0$, particles of species $\sps$ align their direction of motion with that of particles of species $\su$, while for $\chi^{\sps\su} < 0$ they anti-align.
Throughout this work, we set $\chi^{\sa\sa} = \chi^{\spb\spb} = 1$.
We define
$\bar\chi = \tfrac{1}{2}(\chi^{\sa\spb} + \chi^{\spb\sa})$, which quantifies the net tendency for alignment ($\bar\chi > 0$) or anti-alignment ($\bar\chi < 0$), as well as $\Delta \chi = \tfrac{1}{2}(\chi^{\sa\spb} - \chi^{\spb\sa})$, which measures the degree of nonreciprocity.
Interactions are reciprocal when $\Delta\chi = 0$, and nonreciprocal otherwise.

%%%%%%%%%%%%%%%%%%%%%%%%%%%%%%%%%%%
\begin{figure}[t!]
    \centering
    \includegraphics[width=\columnwidth]{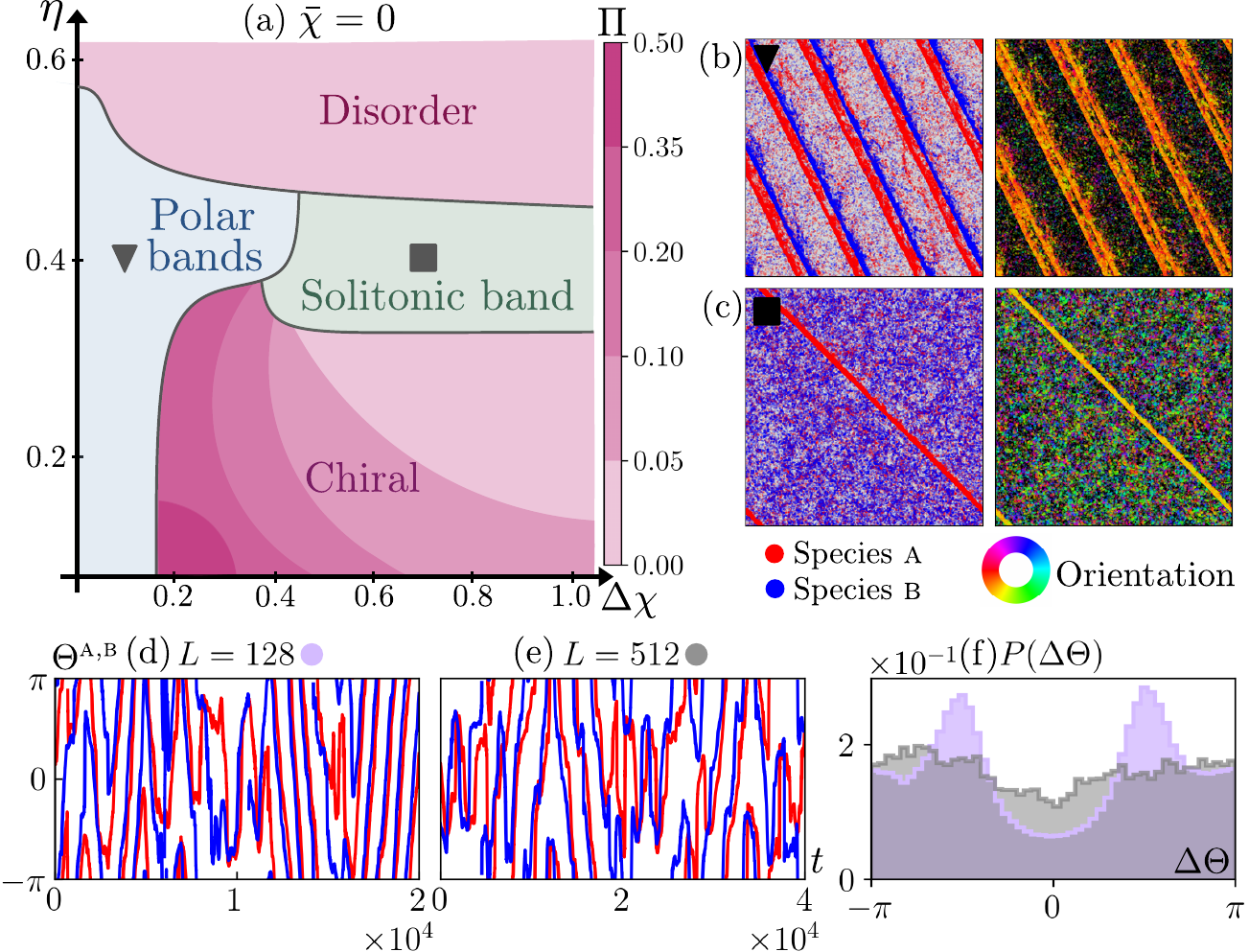}
    \caption{
    Phase behaviour of nonreciprocal flocks in the strong nonreciprocity regime.
    (a) Schematic phase diagram in the ($\Delta \chi,\eta$) plane with $L = 512$.
    The `chiral' region is coloured according to the global polarity averaged over species $\Pi = \tfrac{1}{2}(\Pi^\sa + \Pi^\spb)$.
    See~\cite{supplement} for additional details on the labeling of the phases. 
    (b,c) Representative snapshots of the parallel bands (b) and solitonic band (c) phases.
    Particles are coloured according to species (orientation) in the left (right) panel (caption below),
    the corresponding points in (a) are indicated with solid symbols.
    (d,e) Time series of the global polarity orientations $\Theta^{\sa,\spb}$ for $\Delta\chi = 0.2$ and $\eta = 0.2$ with $L = 128$(d) and $512$(e). The red(blue) curves correspond to species $\sa$($\spb$). 
    (f) Corresponding probability distributions of $\Delta \Theta = \Theta^{\spb} - \Theta^{\sa}$ for $L=128$ (purple) and $512$ (gray).}
    \label{fig:fig1}
\end{figure}
%%%%%%%%%%%%%%%%%%%%%%%%%%%%%%%%%%%

The phase diagram of the model~\eqref{micromodel} in the weakly nonreciprocal regime $|\bar\chi| > |\Delta \chi|$ has been characterized in Ref.~\cite{myin2026}. In this case, the two species mutually align or anti-align, albeit with different strengths, and typically self-organize into various homogeneous phases or spatial patterns that are globally polar, anti-polar, or disordered.
Hereafter, we instead focus on the strongly nonreciprocal regime, $|\Delta \chi| > |\bar\chi|$, where one species aligns with the other while the latter simultaneously anti-aligns.
This antagonistic coupling induces dynamical frustration, which underlies the emergence of rotating phases spontaneously breaking chiral and time-translation symmetries~\cite{Fruchart2021Apr}.

From now on, we fix $v_0 = 1$ and focus on the case $\bar\chi = 0$ (see Supplemental Material~\cite{supplement} for additional data at $\bar\chi \neq 0$, leading to qualitatively similar results).
Equations~\eqref{micromodel} are simulated in periodic square domains of linear size $L$ with equal number densities $\rho^{\sa,\spb} = \rho_0 = \tfrac{1}{2}$.
Using the fact that this setting is symmetric upon exchanging $\sa \leftrightarrow \spb$ and the sign of $\Delta\chi$,
we further restrict our analysis to $\Delta\chi > 0$, so that $\chi^{\sa\spb} > 0$ and $\chi^{\spb\sa} < 0$.

A phase diagram in the noise-nonreciprocity plane is shown in Fig.~\ref{fig:fig1}(a) for systems of size $L = 512$.
On the $\Delta\chi = 0$ line, the two species do not interact and each behaves as an independent flock, 
undergoing a liquid-gas-like transition to collective motion as the noise $\eta$ is decreased~\cite{DADAM}.
As for single-species flocks, the coexistence phase consists of a periodic array of dense, ordered bands propagating in a dilute, disordered background.
For small $\Delta\chi > 0$, the species primarily self-organize into spatially synchronized arrangements of traveling-bands~\footnote{For $\bar\chi = 0$, the two species never form a homogeneous ordered phase. For $\bar\chi > 0$ and $\bar\chi < 0$, polar and anti-polar homogeneous ordered phases, respectively, may be found at sufficiently small $\Delta\chi$, see~\cite{supplement}.}, as shown in Fig.~\ref{fig:fig1}(b).
In each pair of bands, the aligning species (here $\sa$) resides at the front, and the anti-aligning species ($\spb$) at the back.
As discussed in Refs.~\cite{martin2025transition,myin2026}, a similar synchronization mechanism occurs for mutually aligning species and arises from the nonreciprocal nature of the interactions.
%Remarkably, it allows the two species to self-organize into a globally polar pattern even though $\bar\chi = 0$.

For larger values of $\Delta\chi$, the high-noise disordered phase is bordered 
by a new phase [green region in Fig.~\ref{fig:fig1}(a)] in which species $\sa$ forms a localized solitonic band propagating through a globally disordered gas,
as displayed in Fig.~\ref{fig:fig1}(c).
When initializing simulations with multiple bands, for instance by duplicating an existing configuration, 
these structures rapidly merge, indicating that this state should always consists of a single band [see Supplementary Movie (SMov)~1].
For sufficiently large $\Delta\chi$, this band typically reaches densities $\gg \rho_0$, 
such that it remains remarkably stable even when $\Delta\chi \to 1$ where $\chi^{\sa\sa} \simeq \chi^{\sa\spb}$, despite $\sa$ particles continuously interacting with a disordered bath of $\spb$ particles.
A more detailed investigation of this structure will be presented elsewhere.

Further decreasing the noise, the system enters a homogeneous phase in which chiral and time-translation symmetries are spontaneously broken.
In sufficiently small systems and for weak nonreciprocity, the orientations $\Theta^{\sps}(t)$ of both species global polar order,
$\bm \Pi^\sps = \langle \hat{\bm u}[\theta_k^\sps] \rangle_k = \Pi^\sps \hat{\bm u}(\Theta^\sps)$, 
exhibit persistent rotations with a well-defined angular frequency, as reported in Fig.~\ref{fig:fig1}(d) and SMov.~2.
This chiral rotation corresponds to a chasing dynamics of the global polarities, 
which maintain a finite phase shift $\Delta\Theta = \Theta^\spb - \Theta^\sa$.
Rotations are occasionally interrupted by reversals, 
such that clockwise and counterclockwise rotations occur with equal probability, 
as evidenced by the bimodality of the distribution $P(\Delta\Theta)$ shown in Fig.~\ref{fig:fig1}(f).

%%%%%%%%%%%%%%%%%%%%%%%%%%%%%%%%%%%
\begin{figure*}[t!]
    \centering
    \includegraphics[width=\linewidth]{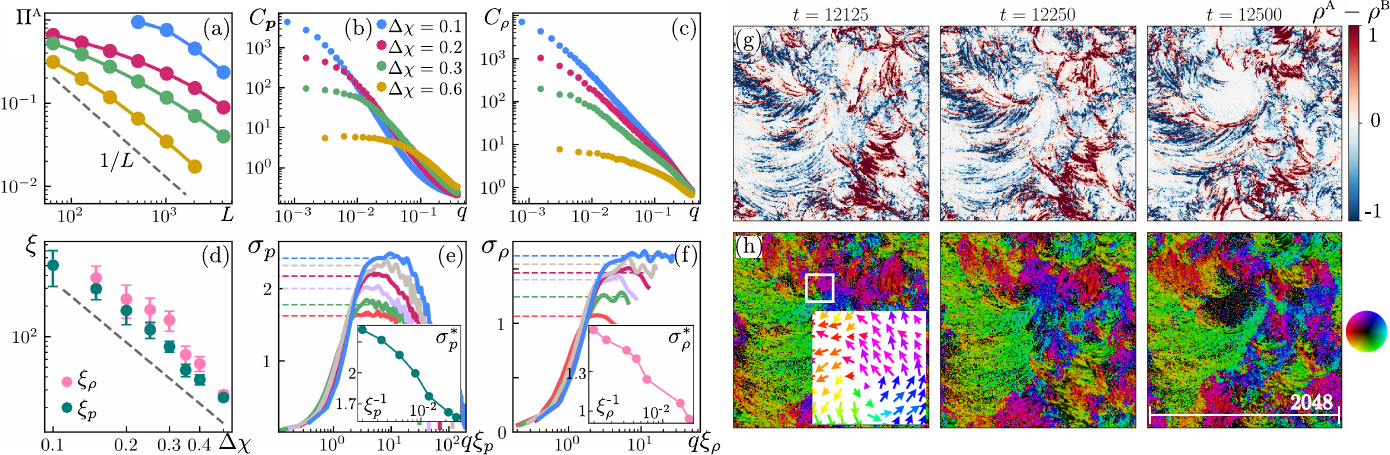}
    \caption{
    Characterization of the nonreciprocity-induced defect-chaos phase ($\eta = 0.2$).
    (a) Time-averaged polar order parameter associated with species $\sa$ as a function of system size for various values of $\Delta\chi$.
    (b,c) Polarity (b) and density fluctuations (c) equal-time correlation function for species $\sa$ 
    radially averaged over the ($q_x,q_y$) plane as a function of the wavenumber $q = \sqrt{q_x^2 + q_y^2}$.
    Legend for (a,b,c) is in (b).
    (d) Correlation lengths extracted from $C_p$ and $C_\rho$ (see text for details).
    The dashed line indicates $\xi \simeq \Delta\chi^{-2}$ as a guide to the eye.
    (e,f) Local exponent associated with $C_p$ (e) and $C_\rho$ (f) (see text) 
    as a function of $q \xi$ for $\Delta \chi = 0.10,0.15,0.20,0.25,0.30,0.35$, from top to bottom.
    Insets show the fitted plateau levels (horizontal dashed lines in main panels) as functions of the correlation length.
    In (b-f), system sizes are $L = 8192$ for $\Delta \chi < 0.2$, $L  = 4096$ for $0.2 \le \Delta \chi <0.4$, and 
    $L  = 2048$ otherwise.
   (g,h) Snapshots showing the nucleation and growth of a disordered bubble whose diameter at $t = 12500$ reaches several hundreds interaction radii for $\Delta \chi = 0.1$.
    (g) shows the relative species density and (h) the orientation of the local polarity of $\sa$ particles, where disordered regions appear black.}
    \label{fig:fig2}
\end{figure*}
%%%%%%%%%%%%%%%%%%%%%%%%%%%%%%%%%%%

However, increasing the system size leads to a rapid deterioration of global order.
Reversal events become increasingly frequent [Fig.~\ref{fig:fig1}(e)], 
leading to the progressive flattening of $P(\Delta\Theta)$,
while the global polarities $\Pi^\sps$ decay rapidly as $L$ increases [Fig.~\ref{fig:fig2}(a)].
As a result, the noise-dependent threshold value of $\Delta\chi$ below which global order is perceptible
becomes lower in larger systems.
For sufficiently small $\Delta\chi$, 
the homogeneous chiral phase is also metastable to the nucleation of polar-bands.
The characteristic nucleation time, however, increases rapidly with system size.
In sufficiently large systems, a homogeneous chiral phase can thus be stabilized for several millions time steps even when $\Delta\chi$ approaches zero.

%%%%%%%%%%%%%%%%%%%%%%%%%%%%%%%%%%%
%     Chiral phase phenomenology
%%%%%%%%%%%%%%%%%%%%%%%%%%%%%%%%%%%

\textit{Breakdown of chiral order and defect-chaos phase.---}
Increasing system size, the chiral phase gradually becomes populated by topological defects
[Fig.~\ref{fig:fig2}(h) and SMov.~3].
The presence of these defects induces a finite correlation length, which is responsible for the rapid decay of the global polar order with increasing system size.
We measured this correlation length by computing the the coarse-grained density
$\rho^\sps(\bm r,t)= \sum_k \delta(\bm r-\bm r_k^\sps(t))$
and polarity
$\bm p^{\sps}(\bm r,t)=\rho^{\sps}(\bm r,t)^{-1}\sum_k \hat{\bm u}[\theta_k^\sps(t)]\delta(\bm r-\bm r_k^\sps(t))$ fields.
In Fourier space, the equal-time two-point correlation functions of both density fluctuations $\rho^\sps(\bm r,t)-\rho_0$
and polarity exhibit a plateau at small wavenumber $q$, as shown in Figs.~\ref{fig:fig2}(b,c) for species $\sa$ (see Fig.~\ref{fig:speciesB} in End Matter for analogous results for $\spb$).
Defining the density and polarity correlation lengths, $\xi_\rho$ and $\xi_p$, from the half-width at half maximum of the $q=0$ peak,
we show in Fig.~\ref{fig:fig2}(d) that both $\xi_{\rho}$ and $\xi_{p}$ grow upon reducing $\Delta\chi$,
with a trend compatible with $\xi_{\rho,p}\sim \Delta\chi^{-2}$.

For weak nonreciprocity, $\xi_p$ becomes sufficiently 
large to unveil an apparent algebraic decay of the polarity correlation function over an intermediate range of wavenumbers
[Fig.~\ref{fig:fig2}(b)].
Remarkably, the associated exponents appear to vary with $\Delta\chi$.
This can be seen most clearly by examining the local exponent of the correlation function $C_p(q)$,
defined as $\sigma_p = -{\rm d}\ln C_p/{\rm d}\ln q$, 
which grows from zero and reaches a plateau $\sigma_p^*$ at intermediate wavenumbers, 
corresponding to a local power law decay $C_p(q) \simeq q^{-\sigma_p^*}$.
As shown in Fig.~\ref{fig:fig2}(e), 
$\sigma_p^*$ increases systematically as $\Delta\chi$ decreases and $\xi_p$ grows,
while it does not show any apparent sign of saturation.
The accuracy of our estimates is limited by the finite correlation length, which restricts the range of the scaling regime.
Within this limitation, we find effective decay exponents in the range $1.5 \lesssim \sigma_p^* \lesssim 2.5$ and obtain comparable values for species $\spb$,
while we expect larger values as $\Delta\chi \to 0$.

Density correlations also exhibit intermediate algebraic decay, albeit with smaller exponents.
Defining $\sigma_\rho^*$ from the plateaus associated with the local exponent of $C_\rho(q)$ [Figs.~\ref{fig:fig2}(c,f)],
we find that that it increases as well when reducing nonreciprocity, and takes values $\sigma_\rho^* \in [1;1.6]$ similar for both species.
To our knowledge, this behavior does not correspond to any known universality class.
In conventional flocks, the advection of density by self-propulsion, in particular,
enforces $C_\rho(q)\propto C_p(q)$, thus fixing $\sigma_p^*=\sigma_\rho^*$~\cite{tonerPRE98,mahaultPRL2019}.
The violation of this relation hence suggests that the presence of strong nonreciprocity qualitatively modifies the large-scale description of the system.

The chiral phase is, however, constantly populated by topological defects, whose dynamics is profoundly affected by the presence of self-propulsion.
As illustrated in Figs.~\ref{fig:fig2}(g,h) and SMov.~3, spiral-like defects act as nucleation sites for propagating circular perturbations.
Since particles at the core of a vortex-like defect predominantly point outward,
once nucleated this structure generates a nearly empty bubble whose radius grows rapidly,
leaving an extended dilute and disordered region in place of the original defect.
Although the depleted region is eventually refilled by incoming particles,
the typical bubble size grows with the correlation length and can reach several hundred interaction ranges for low values of $\Delta\chi$.
We naturally expect the continuous generation of such perturbations
to significantly affect both density and polarity statistics on scales below the correlation length.
We therefore hypothesize that this intrinsically nonlinear mechanism 
may be responsible for the unconventional scaling behavior reported above.

%%%%%%%%%%%%%%%%%%%%%%%%%%%%%%%%%%%
%           PDE results
%%%%%%%%%%%%%%%%%%%%%%%%%%%%%%%%%%%

\begin{figure}[t!]
    \centering
    \includegraphics[width=\columnwidth]{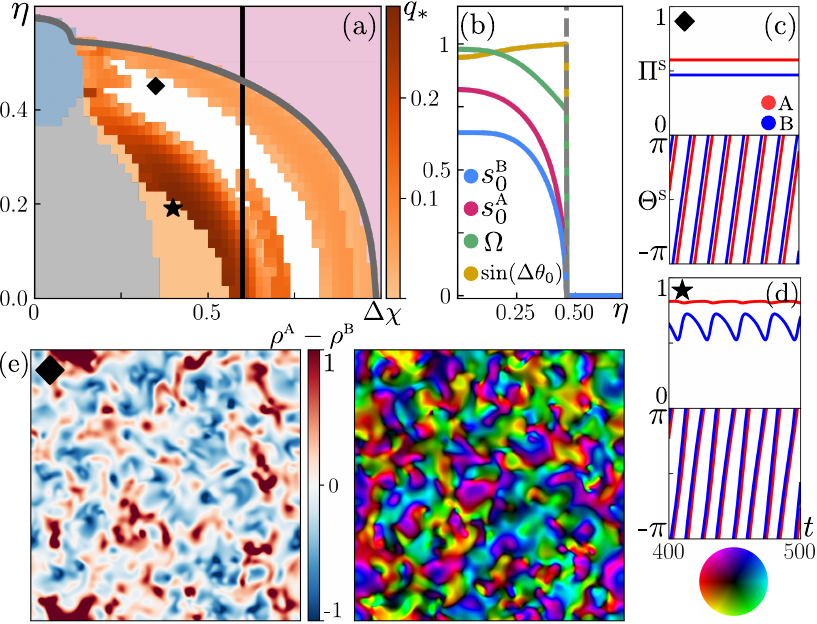}
    \caption{Phenomenology of the continuum model.
    (a) Phase diagram showing regions where
    Eqs.~\eqref{eq_hydro} admit disordered and homogeneous polar solutions in purple and blue, respectively.
     Within the orange domain, homogeneous solutions are associated with global chiral order.
     The color bar indicates the magnitude of the most unstable wavenumber obtained from Floquet stability analysis, while white indicates that the solution is linearly stable.
     In the grey region, Eqs.~\eqref{eq_hydro} exhibit unphysical solutions~\cite{footnoteCM}.
     (b) Homogeneous uniform chiral solution as a function of the noise $\eta$ for $\Delta \chi = 0.6$ (black line in (a)).
     (c,d) Time series of the uniform (c) and modulated (d) chiral solutions.
    Parameters are $(\Delta\chi,\eta) = $
    $(0.35,0.45)$ (c) and $(0.4,0.2)$ (d),
    marked by the black diamond and star in (a), respectively.
    (e) Representative simulation snapshots
    showing the defect-chaos state obtained for the parameters of (c) and $L = 256$ from a random initial condition.
    The left and right panels
    show the relative species density and the orientation of the local momentum of $\sa$ particles (caption on the right), respectively.
    }
    \label{fig:fig3}
\end{figure}

\paragraph{Continuous description.---}
To elucidate the emergence of the defect chaos state, we derived a continuum theory for nonreciprocal flocks by coarse-graining the microscopic model~\eqref{micromodel}.
The derivation follows standard techniques~\cite{bertinPRE2006,Peshkov2014Jun,ChateLectureNotes} and generalizes the one performed in Ref.~\cite{myin2026} for multi-species flocks without chiral order; details will be presented elsewhere.
The resulting theory involves two scalar and two vector fields, corresponding to the coarse-grained densities $\rho^{\sps}$ and momenta $\bm w^{\sps} = \rho^{\sps} \bm p^{\sps}$ of both species $\sps = \sa,\spb$.
The equations are derived perturbatively close to the ordering threshold, marked by the grey line in Fig.~\ref{fig:fig3}(a), as an expansion in gradients and the magnitudes of the momenta.
They read (for $i=x,y$ and using summation over repeated indices)
\begin{subequations}
\label{eq_hydro}
\begin{align} 
    \label{eq_hydro_rho}
    & \partial_t \rho^{\sps} + v_0 \nabla \cdot \bm w^{\sps} = D_\rho^\sps \nabla^2 \rho^\sps, \\
    & \partial_t w_i^{\sps} + \psi_{\su\sv}^{\sps} (\bm w^{\sv}\cdot\nabla) w_i^{\su}
    + \lambda_{\su\sv}^{\sps}(\nabla\cdot\bm w^{\sv}) w_i^{\su}
    + \nu_{\su\sv}^{\sps}\bm w^{\su} \cdot \nabla_i \bm w^{\sv} \nonumber \\
    & + \tilde\psi_{\su\sv}^{\sps} (\bm w^{\sv}\cdot\nabla) w^{\su}_{\perp,i}
    + \tilde\lambda_{\su\sv}^{\sps}(\nabla\cdot\bm w^{\sv}) w^{\su}_{\perp,i}
    + \tilde\nu_{\su\sv}^{\sps} \bm w^{\su} \nabla_{\perp,i} \, \bm w^{\sv} \nonumber \\
    & =  \left[\alpha^{\sps}_{\su}[\rho^\sa,\rho^\spb] 
    - \xi^{\sps}_{\su\sv\sw}(\bm w^{\sv}\cdot \bm w^{\sw}) \right] w_i^{\su} 
    - \tilde\xi^{\sps}_{\su\sv\sw}(\bm w^\sv \cdot \bm w^\sw) w_{\perp,i}^\su \nonumber \\
    \label{eq_hydro_p}
    & - \frac{v_0}{2}\nabla \rho^\sps + D^{\sps}_{\su}\nabla^2 w_i^{\su} + \tilde D^\sps_\su \nabla^2 w_{\perp,i}^\su ,
\end{align}
\end{subequations}
where the subscript $\perp$ denotes vectors rotated by an angle $\pi/2$.

%%%%%%%%%%%%%%%%%%%%%%%%%%%%%%%%%%%
%%%%%%%%%%%%%%%%%%%%%%%%%%%%%%%%%%%

As detailed in~\cite{supplement}, coefficients in Eqs.~\eqref{eq_hydro} depend on the particle-level parameters $\rho_0$, $v_0$, $\eta$, and $\chi^{\sps\su}$ in a complex manner. 
Compared to continuum theories formulated in related settings~\cite{Fruchart2021Apr,kreienkampComPhys2025,myin2026}, 
Eqs.~\eqref{eq_hydro} feature apparent \emph{odd} contributions, including diffusive ($\propto \tilde D^\sps_\su$), advective ($\propto \tilde\psi_{\su\sv}^{\sps}, \tilde\lambda_{\su\sv}^{\sps}, \tilde\nu_{\su\sv}^{\sps}$), and effective potential ($\propto \tilde\xi^{\sps}_{\su\sv\sw}$) terms.
As required by the chiral symmetry of the microscopic dynamics, 
which demands that chiral solutions of Eqs.~\eqref{eq_hydro} must spontaneously break parity symmetry, 
all tilded coefficients are pseudoscalars and proportional to $\sin(\Delta\theta)= (\bm w^\sa \times \bm w^\spb)/|\bm w^\sa||\bm w^\spb|$. 
Whenever $\Delta \chi \gg \bar\chi$,
Eqs.~\eqref{eq_hydro} indeed admit uniform rotating chiral solutions characterized by homogeneous densities $\rho_0^\sps$ and polar orders $\bm w^\sps_0(t) = s_0^\sps \hat{\bm u}(\Omega t + \theta^\sps_0)$ 
rotating at a uniform frequency $\Omega$ and where the two species display a constant phase shift $\Delta\theta_0 = \theta^\spb_0 - \theta^\sa_0$ [Fig.~\ref{fig:fig3}(c)].
Since both $\Omega$ and $\sin\Delta\theta_0$ are generally $\mathcal{O}(1)$, they undergo a discontinuous jump at the ordering threshold [Fig.~\ref{fig:fig3}(b)], such that the odd terms and their even counterparts in Eqs.~\eqref{eq_hydro} are effectively of same order in the expansion.
When $\Delta\chi$ is sufficiently small, on the other hand, rotating chiral solutions do not exist, $\sin\Delta\theta_0=0$, and Eqs.~\eqref{eq_hydro} reduce to the continuum theory derived in~\cite{myin2026}.
Setting $\bar\chi=0$, Eqs.~\eqref{eq_hydro}
also admit homogeneous rotating chiral solutions with weak amplitude modulations at low noises [Fig.~\ref{fig:fig3}(d)],
similar to those reported in Ref.~\cite{Fruchart2021Apr}.
Finally, for sufficiently low $\Delta\chi$ and high noises chiral solutions are absent and Eqs.~\eqref{eq_hydro} show homogeneous polar solutions [blue region in Fig.~\ref{fig:fig3}(a)]~\footnote{Polar solutions of Eqs.~\eqref{eq_hydro} are generally unstable.}.

We performed the Floquet stability analysis of the uniform rotating chiral solution $\rho_0^\sps,\bm w^\sps_0(t)$ (details in~\cite{supplement}).
As summarized in Fig.~\ref{fig:fig3}(a), this solution is unstable for most parameters where it exists, with a linearly stable pocket persisting only at intermediate $\Delta\chi$ [white region in Fig.~\ref{fig:fig3}(a)].
Numerical integration of the full nonlinear equations~\eqref{eq_hydro} shows that this instability leads to a chaotic regime characterized by the proliferation of topological defects and strong density modulations, reminiscent of active turbulence~\cite{alertAnnrev2022} [Fig.~\ref{fig:fig3}(e) and SMov.~4].
Modulated chiral solutions are also unstable, leading to the same outcome.
In addition, simulations initialized with random initial conditions in the region where the uniform rotating chiral solution is linearly stable converge again to the defect-chaos state, 
implying that chiral order is only metastable.
These results therefore confirm the scenario evidenced by microscopic simulations: long-range chiral order is generically destroyed by the nucleation and proliferation of topological defects, inducing a finite correlation length.

%%%%%%%%%%%%%%%%%%%%%%%%%%%%%%%%%%%
%           Conclusion
%%%%%%%%%%%%%%%%%%%%%%%%%%%%%%%%%%%

\textit{Discussion---}
We have shown that, in the strong nonreciprocity regime, multispecies flocks can self-organize into globally polar states through the formation of inhomogeneous traveling band structures.
Nevertheless, such systems cannot sustain large-scale chiral order.
We demonstrated that chiral order is ultimately destroyed by the proliferation of topological defects, which generate a finite correlation length and cause the rapid decay of global order.
These results are further supported by the analysis of a coarse-grained hydrodynamic description, 
which reveals that chiral solutions are generically unstable or, at best, metastable.
In all cases, the dynamics converges to a defect-chaos state dominated by strong spatiotemporal density and polarity fluctuations.

Our numerical simulations further reveal that nonreciprocal flocks exhibit unconventional scaling behavior characterized by nonuniversal exponents, in stark contrast with theoretical expectations based on KPZ roughening, which predict a universal value $\sigma_p^* \simeq 2.8$~\cite{canetPRL2010,klossPRE2012}. 
We trace this discrepancy to the peculiar role of topological defects, which---through the intrinsic coupling between density and polarity---act as persistent sources of nonlinear fluctuations at scales below the correlation length. 
Related feedback mechanisms have been discussed in other classes of polar active matter, where they modify the order of phase transitions~\cite{mahaultPRL2018} and affect the coarsening dynamics~\cite{chardacPRX2021}. 
Understanding how such density-polarity coupling controls the large-scale behavior of nonreciprocal flocks constitutes an open challenge for future work.\\

\acknowledgements
C.M. acknowledges funding from the International Max Planck Research School (IMPRS) for the Physics of Biological and Complex Systems, and thanks Ramin Golestanian and the Living Matter Department at MPI-DS for supporting her to visit the Charles Coulomb laboratory in Montpellier, and the laboratory for hosting her during part of this work.

\bibliographystyle{apsrev4-2}
\bibliography{sample.bib}

\onecolumngrid

\clearpage

\vspace{12pt}
\noindent\hrulefill \hspace{24pt} {\bf End Matter} \hspace{24pt} \hrulefill
\vspace{12pt}

%\twocolumngrid

\renewcommand \thefigure{A\arabic{figure}}
\setcounter{figure}{0}
\setcounter{equation}{0}
\renewcommand{\theequation}{A\arabic{equation}}
% \hypertarget{appA}{\textit{Appendix A: ---}}

%%%%%%%%%%%%%%%%%%%%%%%%%%%%%%%%%%%
\begin{figure*}[h!]
    \centering
    \includegraphics[width=\linewidth]{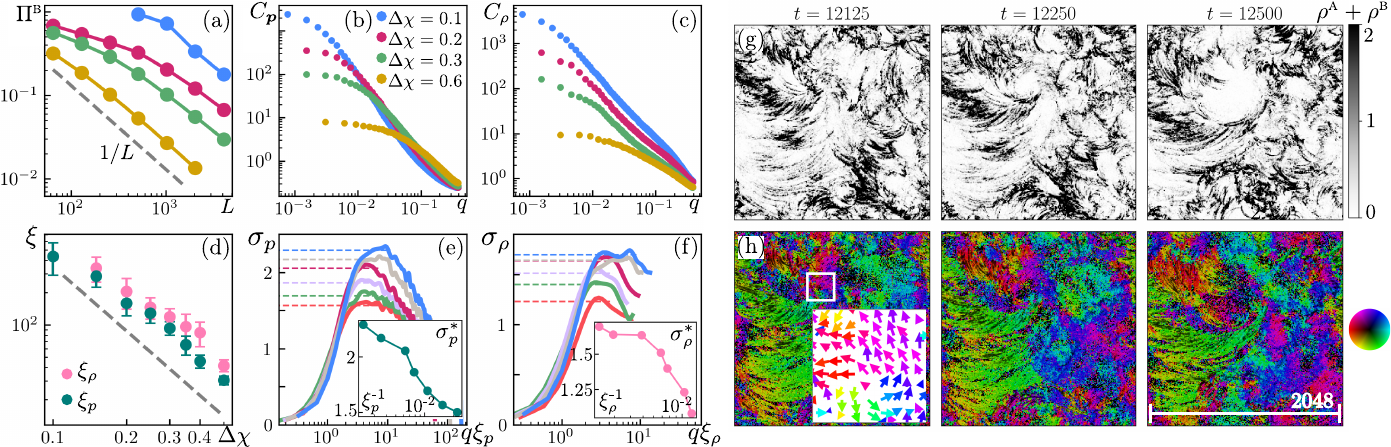}
    \caption{This figure corresponds to Fig.~\ref{fig:fig2} of the main text, but with global order, correlation functions and correlation length extracted from species $\spb$ with $\eta = 0.2$.
    (a) Time-averaged polar order parameter associated with species $\spb$ as a function of system size for various values of $\Delta\chi$.
    (b,c) Polarity (b) and density fluctuations (c) equal-time correlation function for species $\spb$ 
    radially averaged over the ($q_x,q_y$) plane as a function of the wavenumber $q = \sqrt{q_x^2 + q_y^2}$.
    Legend for (a,b,c) is in (b).
    (d) Correlation lengths extracted from $C_p$ and $C_\rho$ (see main text for details).
    The dashed line indicates $\xi \simeq \Delta\chi^{-2}$ as a guide to the eye.
    (e,f) Local exponent associated with $C_p$ (e) and $C_\rho$ (f) (see text) 
    as a function of $q \xi$ for $\Delta \chi = 0.10,0.15,0.20,0.25,0.30,0.35$, from top to bottom.
    Insets show the fitted plateau levels (horizontal dashed lines in main panels) as functions of the correlation length.
    In (b-f), system sizes are $L = 8192$ for $\Delta \chi < 0.2$, $L  = 4096$ for $0.2 \le \Delta \chi <0.4$, and 
    $L  = 2048$ otherwise.
   (g,h) Snapshots showing the nucleation and growth of a disordered bubble whose diameter at $t = 12500$ reaches several hundreds interaction radii for $\Delta \chi = 0.1$.
    (g) shows the total species density and (h) the orientation of the local polarity of $\spb$ particles, where disordered regions appear black.}
    \label{fig:speciesB}
\end{figure*}
%%%%%%%%%%%%%%%%%%%%%%%%%%%%%%%%%%%

\clearpage

\onecolumngrid 
\normalfont
\renewcommand \thefigure{S\arabic{figure}}
\setcounter{figure}{0}
\setcounter{equation}{0}
\renewcommand{\theequation}{S\arabic{equation}}
\begin{center}
\textbf{\Large Supplemental Material: Breakdown of Emergent Chiral Order and Defect Chaos in Nonreciprocal Flocks}
\end{center}
\vspace{2ex}
In this document we provide additional details on the construction of the phase diagram shown in Fig 1 of the main text, and we demonstrate that the results reported there are robust to the introduction of a small symmetric component in the interactions. We further expand on the continuum theory (Eqs. (2) in the main text). Specifically, we give the explicit expressions of the coefficients, discuss the homogeneous solutions and the procedure used to determine their stability, and present technical details regarding the numerical evaluation of the equations.  Finally, we include a description of the accompanying supplemental movies. 

\section{Agent-based simulations}
\subsection{Phase diagrams}

\begin{figure}[h]
    \centering
    \includegraphics[width=0.35\linewidth]{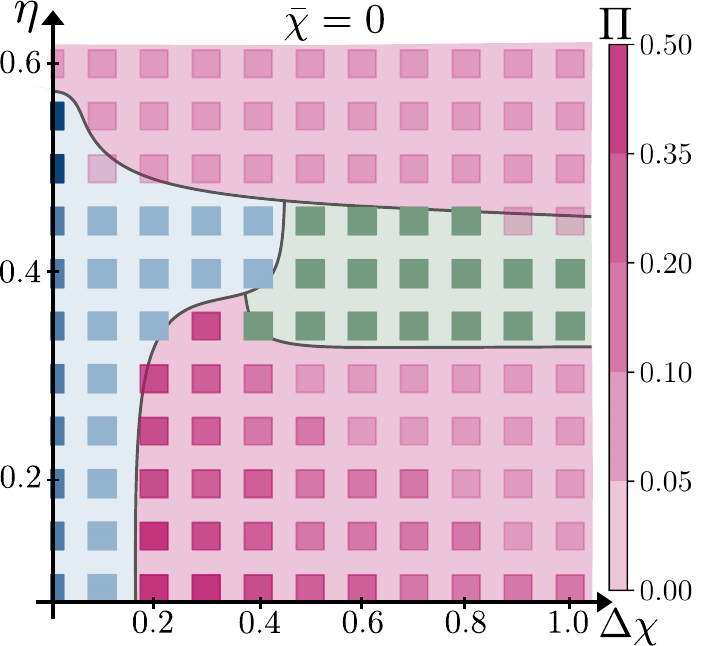}
    \caption{Phase diagram in the $(\Delta\chi,\eta)$ plane for $L=512$. Each square corresponds to an independent simulation and is color-coded according to the observed steady-state phase: polar order (blue)---subdivided into homogeneous flocking (blue), independent Vicsek bands (dark blue), and parallel polar bands (light blue)--- solitonic band (green) and chiral-disordered (pink). Shades of pink correspond to the mean polar order $\Pi= \tfrac{1}{2}\big( \bar \Pi^{\sa}  +  \bar \Pi^{\spb}\big)$ (legend on the right). The classification criteria are provided in the text.}
    \label{fig:figS1}
\end{figure}

The finite-size phase diagram shown in Fig.~1(a) of the main text is constructed from a set of simulations systematically spanning the $(\Delta\chi,\eta)$ plane with $\bar\chi = 0$ and $L = 512$.
Figure~\ref{fig:figS1} displays the data used determine the phase behaviour of the model, which each square corresponding to a single run.
Each point is labelled according to the time-averaged global polar order parameters $\bar\Pi^{\sa,\spb} = \langle \Pi^{\sa,\spb} \rangle_t$.
and their corresponding orientation $\Delta\bar\Theta = \langle |\Theta^{\sa} - \Theta^{\spb}|\rangle_t$.
These quantities are averaged over typically $8 \times 10^5$ simulation steps after the system reaches steady-state.
Our classification criteria read:
\begin{itemize}
    \item {\bf Polar order (blue)}
    \begin{itemize}
        \item (Independent) homogeneous flocking: $\bar \Pi^\sa  > 0.8$, $ \bar \Pi^\spb > 0.8$ and $\Delta\bar\Theta > 0.05$ (blue)
        \item (Independent) Vicsek bands: $ \bar \Pi^\sa > 0.65$, $\bar \Pi^\spb > 0.65$ and $\Delta\bar\Theta > 0.05$ (dark blue)
        \item Parallel synchronized bands: $\bar \Pi^\sa > 0.6$, $ \bar \Pi^\spb > 0.6$ and $\Delta\bar\Theta \le 0.05$ (light blue)
    \end{itemize} 
    \item {\bf Solitonic band (green)}: $\bar \Pi^\sa > 0.4$, $ \bar \Pi^\spb < 0.1$ 
    \item {\bf Chiral and disorder (pink)}: if none of the above is true. 
    The color shading is determined by averaging polar order over both species: $\Pi = \tfrac{1}{2}( \bar \Pi^\sa  +\bar \Pi^\spb)$
\end{itemize}
These are meant, in particular, to distinguish between the synchronized parallel band patterns (Fig.~1(b) of the main text), where both species exhibit aligned global polar order, and the solitonic band (Fig.~1(c) of the main text), for which only species $\sa$ is globally ordered. 
As shown in the main text, since global chiral order is associated with a system size-independent correlation length, the low noise, large nonreciprocity regime is eventually disordered. 
To mark the distinction with the disordered phase found at high noises, for which the correlation length typically takes smaller values, the purple points in Fig.~\ref{fig:figS1} are colored according the time and species-averaged global order $\Pi$.
The phase boundaries in Fig.~1(a) are then determined by separating points with different colors.
Note that Vicsek bands and homogeneous flocking of both species only exist for $\Delta\chi = 0$ and are thus not represented in Fig.~1(a).

The parameters corresponding to the representative snapshots in Fig. 1(b,c) of the main text are $\rho_0 = \frac{1}{2}$, $v_0 = 1$, $\eta = 0.4$ and $\bar\chi = 0$. Panel (b) corresponds to $\Delta\chi = 0.1$, and panel (c) to $\Delta\chi = 0.7$. 

\pagebreak

\subsection{Systems with global alignment and anti-alignment ($\bar\chi \ne 0$)}

\begin{figure}[!h]
    \centering
\includegraphics[width=0.8\linewidth]{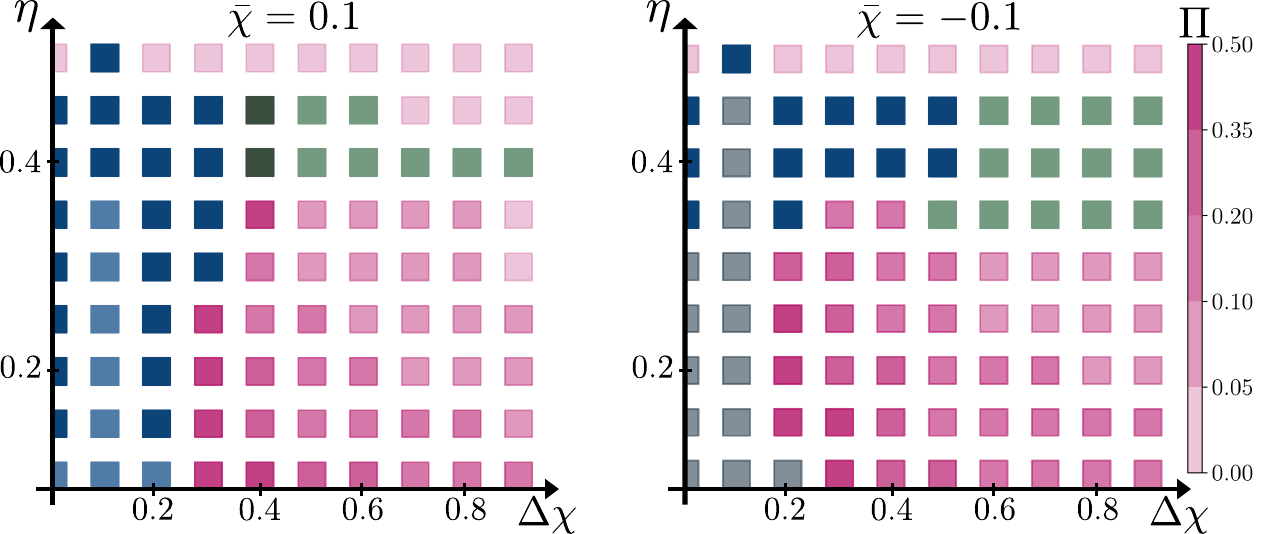}
    \caption{Phase diagrams in the $(\Delta\chi,\eta)$ plane for $L=256$, $\bar\chi = 0.1$ (left) and $\bar\chi = -0.1$ (right). Each square corresponds to an independent simulation and is color-coded according to its steady-state phase: homogeneous flocking (blue), homogeneous anti-flocking (grey), parallel bands (dark blue), solitonic band (green), and chiral-disordered (pink). Shades of pink correspond to the mean polar order $\Pi= \tfrac{1}{2}\big( \bar \Pi^{\sa}  +  \bar \Pi^{\spb}\big)$ (legend on the right). The classification criteria are provided in the text.}
    \label{fig:figs2}
\end{figure}

The results presented in the main text correspond to purely antisymmetric inter-species alignment: $\bar\chi =0$. 
Here, we show that the phase behaviour of the model is not qualitatively modified by introducing a small symmetric component to the interactions, provided that $|\Delta\chi| > |\bar\chi|$.
We present in Fig.~\ref{fig:figs2} phase-diagrams similar to the one of Fig.~\ref{fig:figS1} for $\bar\chi = 0.1$ and $\bar\chi = -0.1$. 
The main change as compared to $\bar\chi = 0$ case is that the system can now exhibit homogeneous polar ($\bar\chi > 0$) or anti-polar ($\bar\chi < 0$) phases at low noises and for $\Delta\chi \lesssim \bar\chi$.
The $\Delta\chi < \bar\chi$ regime has been thoroughly characterized in Ref.~\cite{myin2026}, based on which we also expect to observe inhomogeneous pattern with demixing of the species considering larger system sizes.
For $\Delta\chi > \bar\chi$, on the other hand, we observe a phenomenology similar to in the $\bar\chi = 0$ case regardless of the sign of $\bar\chi$.
In particular, decreasing the noise the system undergoes a transition from a disordered phase to a parallel bands or solitonic band phases for low and high $\Delta\chi$, respectively.
Remarkably, and as already reported in Ref.~\cite{myin2026}, noreciprocity promotes the emergence of polar patterns even when $\bar\chi < 0$ and the inter-species interactions are globally anti-aligning.

\section{The continuum model}

We will present a detailed derivation of the continuum model in a separate publication~\cite{myin2027}. 
The derivation follows standard techniques~\cite{ChateLectureNotes} and is largely based on the one of~\cite{myin2026} for multi-species flocks in the low nonreciprocity regime.
For convenience, we recall the expression of the continuum theory:
\begin{subequations}
\label{SM_eq_hydro}
\begin{align} 
    \label{SM_eq_hydro_rho}
    & \partial_t \rho^{\sps} + v_0 \nabla \cdot \bm w^{\sps} = D_\rho^\sps \nabla^2 \rho^\sps, \\
    & \partial_t w_i^{\sps} + \psi_{\su\sv}^{\sps} (\bm w^{\sv}\cdot\nabla) w_i^{\su}
    + \lambda_{\su\sv}^{\sps}(\nabla\cdot\bm w^{\sv}) w_i^{\su}
    + \nu_{\su\sv}^{\sps}\bm w^{\su} \cdot \nabla_i \bm w^{\sv} \nonumber
    + \tilde\psi_{\su\sv}^{\sps} (\bm w^{\sv}\cdot\nabla) w^{\su}_{\perp,i}
    + \tilde\lambda_{\su\sv}^{\sps}(\nabla\cdot\bm w^{\sv}) w^{\su}_{\perp,i}
    + \tilde\nu_{\su\sv}^{\sps} \bm w^{\su} \nabla_{\perp,i} \, \bm w^{\sv} \nonumber \\
    \label{SM_eq_hydro_p}
    & =  \left[\alpha^{\sps}_{\su}[\rho^\sa,\rho^\spb] 
    - \xi^{\sps}_{\su\sv\sw}(\bm w^{\sv}\cdot \bm w^{\sw}) \right] w_i^{\su} 
    - \tilde\xi^{\sps}_{\su\sv\sw}(\bm w^\sv \cdot \bm w^\sw) w_{\perp,i}^\su
    - \frac{v_0}{2}\nabla \rho^\sps + D^{\sps}_{\su}\nabla^2 w_i^{\su} + \tilde D^\sps_\su \nabla^2 w_{\perp,i}^\su ,
\end{align}
\end{subequations}
where the subscript $\perp$ denotes vectors rotated by an angle $\pi/2$, e.g.\ $\bm w^\sps_\perp=\hat{\bm z}\times\bm w^\sps$, with $\hat{\bm z}$ the unit vector orthogonal to the plane.

As a result of the coarse-graining procedure, we can express all the coefficients in Eqs.~\eqref{SM_eq_hydro} in terms of the microscopic model parameters. 
In this section, we provide the corresponding expressions of the coefficients, as well as details on the homogeneous solutions of the continuum model, their stability, and technical details about numerical integration of the full equations.

\subsection{Coefficients}

In addition to $v_0$ and $D^\sps_\rho$,
which we fix to $1$ and $D_0 = 0.5$, respectively,
the coefficients appearing in Eqs.~\eqref{SM_eq_hydro_p} can be expressed in terms of mode-coupling coefficients which are given by the following integral~\cite{myin2026}:
\begin{equation} \label{app_intI}
    I_{kq}^{\sps \su} \equiv \frac{1}{2\pi}\int_0^{2\pi}{\rm d}\phi\,
    K(\phi) \cos\left[-i q \phi + i k H^{\sps \su}(\phi)\right].
\end{equation}
where $K(\phi_2-\phi_1) = 2r_0 v_0 \left|\hat{\bm u}(\phi_2) - \hat{\bm u}(\phi_1)\right|$ 
and 
$H^{\sps \su}(\phi_2 - \phi_1) = \,{\rm arg}[ 1 + \chi^{\sps \su}e^{i(\phi_2-\phi_1)}]$.
Here, $r_0 = 1$ corresponds to the interaction radius and $v_0$ to the self-propulsion speed of the particles.
Defining $J_{kq}^{\sps \su} = P_k I_{kq}^{\sps \su} - I_{0 q}^{\sps \su}$ with 
$P_k = \exp[-k^2 \eta^2/2]$ the distribution of the angular noise with strength $\eta$, here assumed Gaussian for simplicity,
the coefficients of Eqs.~\eqref{SM_eq_hydro} can be expressed as follows:
\begin{align}
    \alpha^{\sps}_{\su}[\rho^\sa,\rho^\spb] & = \left[ P_k - 1 + \sum_{\spt} J_{10}^{\sps\spt} \rho^{\spt} \right]\delta^{\sps}_{\su} +  J_{11}^{\sps\su} \rho^{\sps} &
    \barbelow{\xi}^{\sps}_{\su\sv\sw} & = 
    \frac{1}{2}\left[ \barbelow{\beta}^{\sps}_{\sv\su\sw}
    + \barbelow{\beta}^{\sps}_{\sw\su\sv}
    + \barbelow{\beta}^{\sps}_{\sw\sv\su}
    + \barbelow{\beta}^{\sps}_{\sv\sw\su}
    - \barbelow{\beta}^{\sps}_{\su\sv\sw}
    - \barbelow{\beta}^{\sps}_{\su\sw\sv}
    \right], \nonumber \\
    \label{coeffs_SM}
    \barbelow{\psi}_{\su\sv}^{\sps} & = \barbelow{\kappa}_{1,\su\sv}^{\sps} + \barbelow{\kappa}_{1,\sv\su}^{\sps} + \barbelow{\kappa}_{2,\sv\su}^{\sps}, &
    \barbelow{\lambda}_{\su\sv}^{\sps} & = \barbelow{\kappa}_{1,\su\sv}^{\sps} + \barbelow{\kappa}_{1,\sv\su}^{\sps} - \barbelow{\kappa}_{2,\su\sv}^{\sps}, \\
    \barbelow{\nu}_{\su\sv}^{\sps} & = 
    \barbelow{\kappa}_{2,\su\sv}^{\sps} -
    \barbelow{\kappa}_{1,\su\sv}^{\sps} -
    \barbelow{\kappa}_{1,\sv\su}^{\sps}, &
    \barbelow{D}^{\sps}_{\su} & = D_0 \delta^{\sps}_{\su} -\frac{v_0^2}{4}\barbelow{\gamma}^{\sps}_{\su}, \nonumber
\end{align}
where $\delta^{\sps}_{\su}$ is the Kronecker-delta symbol, and we use the underline to indicate when the coefficients are complex. 
The non-tilded and tilded coefficients in Eqs.~\eqref{SM_eq_hydro} then correspond to the real and imaginary part of the coefficients in~\eqref{coeffs_SM}, respectively.

The $\barbelow{\beta}$, $\barbelow{\gamma}$ and $\barbelow{\kappa}$ coefficients are expressed as
(summation over repeated indices is \emph{not} implied)
\begin{align*}
    \barbelow{\beta}^{\sps}_{\su \sv \sw}  & = \barbelow{\gamma}^{\sps}_{\sv}
    J_{1-1,0}^{\sps \su}J^{\sv\sw}_{21}
     + \delta_{\sps\su}\sum_{\spt} \barbelow{\gamma}^{\spt}_{\sv} J_{12,0}^{\sps \spt}J^{\sv\sw}_{21}, &
    \barbelow{\kappa}_{1,\su\sv}^{\sps} &= -\frac{v_0}{2}\barbelow{\gamma}^{\sps}_{\su} J^{\su\sv}_{21}, &
    \barbelow{\kappa}_{2, \su\sv}^{\sps}  & = - \frac{v_0}{2}J_{1-1,0}^{\sps \su}\barbelow{\gamma}^{\sps}_{\sv} - \frac{v_0}{2}\delta_{\sps \su}\sum_{\spt} J_{12,0}^{\sps \spt}\barbelow{\gamma}^{\spt}_{\sv}, 
\end{align*}
where $\barbelow{\gamma}^{\sps}_{\su}$ is defined as the inverse of the matrix
\begin{equation}
    \barbelow{M}^{\sps}_{\su} = [P_k - 1 
+ (\sum_{\sv} J_{k0}^{\sps \sv}\rho_0^{\sv}) - 2 i \Omega]\delta^{\sps}_{\su} + J_{kk}^{\sps \su} \rho_0^{\sps}.
\end{equation}
Although the mode coupling coefficients~\eqref{app_intI} remain strictly real as a result of the chiral symmetry of the microscopic dynamics,
the continuum model involves imaginary coefficients eventually leading to odd terms. 
This is because the continuum equations are formally derived at the onset of a Hopf bifurcation associated with a global angular frequency $\Omega$. 
In practice, we determine the value of $\Omega$ by computing the homogeneous rotating solution of Eqs.~\eqref{SM_eq_hydro} self-consistently (see Sec.~\ref{SM_Homogeneous_solutions}).

However, treating $\Omega$ as a constant parameter in Eqs.~\eqref{SM_eq_hydro} causes the odd terms to explicitly break chiral symmetry.
To account for spontaneous chiral symmetry breaking,
we note that at the onset of chiral order $\Omega$ can generally be expressed as~\cite{myin2027}
\begin{equation*}
    \Omega = \sin[\theta^\spb_0 - \theta^\sa_0] \tilde{\Omega} = \frac{\bm w_0^\sa \times \bm w_0^\spb}{s_0^\sa s_0^\spb} \tilde{\Omega},
\end{equation*}
where $\tilde{\Omega}$ is a positive number and $w_0^{\sa,\spb} = s_0^{\sa,\spb}\hat{\bm u}(\theta_0^{\sa,\spb})$ are the homogeneous rotating solutions of Eqs.~\eqref{SM_eq_hydro}.
Hence, the sign of $\Omega$, which determines the direction of rotation of the solution, is given by the phase shift between the momenta.
To restore the parity symmetry of the continuum model, we thus define 
\begin{equation}\label{new_def_omega}
    \Omega(\bm r,t) = \sin[\theta^\spb(\bm r,t) - \theta^\sa(\bm r,t)] \tilde{\Omega} = \frac{\bm w^\sa \times \bm w^\spb}{|\bm w^\sa| |\bm w^\spb|} \tilde{\Omega},
\end{equation}
where $\theta^{\sa,\spb}(\bm r,t)$ denote the \textit{local} orientations of the momenta.

\subsection{Homogeneous solutions}
\label{SM_Homogeneous_solutions}

To calculate the homogeneous solutions of Eqs.~\eqref{SM_eq_hydro} and study their stability, 
it is convenient to express the momenta as $\bm w^\sps = s^\sps \hat{\bm u}(\theta^\sps)$ 
and rewrite the continuum equations in terms of their magnitudes $s^\sps$ and orientations $\theta^\sps$. 
In compact form, the resulting equations read
\begin{subequations}
\label{SM_eq_hydro_stheta}
\begin{align} \label{SM_eq_hydro_stheta_rho}
    \partial_t \rho^{\sps} & + v_0 \hat{\bm u}^\sps \cdot \nabla s^\sps +  v_0 s^\sps \hat{\bm u}^\sps_\perp \cdot \nabla \theta^\sps  = D_\rho^\sps \nabla^2 \rho^\sps, \\
    %%%%%%%%%%%%%%%%%%%%%%%%%%%%%%
    \partial_t s^\sps &= -\frac{v_0}{2}(\hat{\bm u}^\sps \cdot\nabla) \rho^{\sps} + \sum_{\su}
    \left(F^\sps_\su\cos(\Delta\theta_{\sps\su}) + \Tilde{F}^\sps_\su \sin(\Delta\theta_{\sps\su})\right)s^{\su} \nonumber \\
    &  + \sum_\su D^{\sps}_{\su}\left[ 
    \cos(\Delta\theta_{\sps\su})\left( 
    \nabla^2 s^\su - s^\su |\nabla\theta^\su|^2
    \right)
    + \sin(\Delta\theta_{\sps\su})\left( 
    2(\nabla s^\su)\cdot(\nabla \theta^\su) + s^\su \nabla^2 \theta^\su
    \right)
    \right], \nonumber \\
    & + \sum_\su \Tilde{D}^{\sps}_{\su}\left[ 
    \sin(\Delta\theta_{\sps\su})\left( 
    \nabla^2 s^\su - s^\su |\nabla\theta^\su|^2
    \right)
    - \cos(\Delta\theta_{\sps\su})\left( 
    2(\nabla s^\su)\cdot(\nabla \theta^\su) + s^\su \nabla^2 \theta^\su
    \right)
    \right] \nonumber \\
    \label{SM_eq_hydro_stheta_rho}
    & - \sum_{\su} \left[ \bm K_\su^\sps \cdot \nabla s^\su 
    + s^\su \bm L_\su^\sps \cdot \nabla\theta^\su \right] - \sum_{\su} \left[ \Bar{\bm K}_\su^\sps \cdot \nabla_\perp s^\su 
    + s^\su \Bar{\bm L}_\su^\sps \cdot \nabla_\perp\theta^\su \right],\\
    %%%%%%%%%%%%%%%%%%%%%%%%%%%%%%
    s^\sps \partial_t \theta^\sps &= -\frac{v_0}{2}(\hat{\bm u}^\sps_\perp \cdot \nabla) \rho^{\sps}
    + \sum_{\su} \left[-F^\sps_\su \sin(\Delta\theta_{\sps\su}) 
    + \Tilde{F}^\sps_\su \cos(\Delta\theta_{\sps\su})\right]s^\su \nonumber \\ 
    & + \sum_\su D^{\sps}_{\su}\left[ 
    -\sin(\Delta\theta_{\sps\su})\left( 
    \nabla^2 s^\su - s^\su |\nabla\theta^\su|^2
    \right)
    + \cos(\Delta\theta_{\sps\su})\left( 
    2(\nabla s^\su)\cdot(\nabla \theta^\su) + s^\su \nabla^2\theta^\su
    \right)
    \right] \nonumber \\
    & + \sum_\su \Tilde{D}^{\sps}_{\su}\left[ 
    \cos(\Delta\theta_{\sps\su})\left( 
    \nabla^2 s^\su - s^\su |\nabla\theta^\su|^2
    \right)
    + \sin(\Delta\theta_{\sps\su})\left( 
    2(\nabla s^\su)\cdot(\nabla \theta^\su) + s^\su \nabla^2\theta^\su
    \right)
    \right] \nonumber \\
    \label{SM_eq_hydro_stheta_theta}
    &  - \sum_{\su} \left[\bm M_\su^\sps \cdot \nabla s^\su 
    + s^\su \bm N_\su^\sps \cdot \nabla\theta^\su \right] - \sum_{\su} \left[\Bar{\bm M}_\su^\sps \cdot \nabla_\perp s^\su 
    + s^\su \Bar{\bm N}_\su^\sps \cdot \nabla_\perp \theta^\su \right]
\end{align}
\end{subequations}
where $\hat{\bm u}^\sps = \hat{\bm u}(\theta^\sps)$,  $F^\sps_\su \equiv \alpha^{\sps}_{\su} - \sum_{\sv,\sw}\xi^{\sps}_{\su\sv\sw}s^\sv s^\sw \cos(\Delta\theta_{\sv\sw})$,  
$\Tilde{F}^\sps_\su \equiv  -\sum_{\sv,\sw}\Tilde{\xi}^{\sps}_{\su\sv\sw} s^\sv s^\sw \cos(\Delta\theta_{\sv\sw})$, $\Delta\theta_{\su\sv} \equiv \theta^\su - \theta^\sv$, and
\begin{align*}
    \bm K_\su^\sps &\equiv 
    \cos(\Delta\theta_{\sps\su}) \sum_\sv \psi_{\su\sv}^{\sps} s^\sv \hat{\bm u}^\sv
    + \hat{\bm u}^\su \sum_\sv \lambda_{\sv\su}^{\sps}
    s^\sv \cos(\Delta\theta_{\sps\sv})
    + \hat{\bm u}^\sps \sum_\sv \nu_{\sv\su}^{\sps}
    s^\sv \cos(\Delta\theta_{\su\sv}) \\
    & +  \sin(\Delta\theta_{\sps\su}) \sum_\sv \Tilde{\psi}_{\su\sv}^{\sps} s^\sv \hat{\bm u}^\sv
    + \hat{\bm u}^\su \sum_\sv \Tilde{\lambda}_{\sv\su}^{\sps}
    s^\sv \sin(\Delta\theta_{\sps\sv}) \\
    %%%%%%%%%%%%%%%%%%%%%%%%
    \bm L_\su^\sps & \equiv 
    \sin(\Delta\theta_{\sps\su}) \sum_\sv \psi_{\su\sv}^{\sps} s^\sv \hat{\bm u}^\sv
    + \hat{\bm u}^\su_\perp \sum_\sv \lambda_{\sv\su}^{\sps} 
    s^\sv \cos(\Delta\theta_{\sps\sv})
    - \hat{\bm u}^\sps \sum_\sv \nu_{\sv\su}^{\sps} 
    s^\sv \sin(\Delta\theta_{\su\sv})\\
    & - \cos(\Delta\theta_{\sps\su}) \sum_\sv \Tilde{\psi}_{\su\sv}^{\sps} s^\sv \hat{\bm u}^\sv
    + \hat{\bm u}^\su_\perp \sum_\sv \Tilde{\lambda}_{\sv\su}^{\sps} 
    s^\sv \sin(\Delta\theta_{\sps\sv}) \\
    \Bar{\bm K}_\su^\sps & \equiv 
    \hat{\bm u}^\sps \sum_\sv \Tilde{\nu}_{\sv\su}^{\sps} 
    s^\sv \cos(\Delta\theta_{\su\sv}), \\
    \Bar{\bm L}_\su^\sps & \equiv 
    - \hat{\bm u}^\sps \sum_\sv \Tilde{\nu}_{\sv\su}^{\sps} 
    s^\sv \sin(\Delta\theta_{\su\sv}), \\
    \bm M_\su^\sps &\equiv 
    -\sin(\Delta\theta_{\sps\su}) \sum_\sv \psi_{\su\sv}^{\sps} s^\sv \hat{\bm u}^\sv
    -\hat{\bm u}^\su \sum_\sv \lambda_{\sv\su}^{\sps}
    s^\sv \sin(\Delta\theta_{\sps\sv})
    + \hat{\bm u}^\sps_\perp \sum_\sv \nu_{\sv\su}^{\sps}
    s^\sv \cos(\Delta\theta_{\su\sv}) \\
    & +  \cos(\Delta\theta_{\sps\su}) \sum_\sv \Tilde{\psi}_{\su\sv}^{\sps} s^\sv \hat{\bm u}^\sv
    + \hat{\bm u}^\su \sum_\sv \Tilde{\lambda}_{\sv\su}^{\sps}
    s^\sv \cos(\Delta\theta_{\sps\sv}) \\
    \bm N_\su^\sps & \equiv 
    \cos(\Delta\theta_{\sps\su}) \sum_\sv \psi_{\su\sv}^{\sps} s^\sv \hat{\bm u}^\sv
    - \hat{\bm u}^\su_\perp \sum_\sv \lambda_{\sv\su}^{\sps}
    s^\sv \sin(\Delta\theta_{\sps\sv})
    - \hat{\bm u}^\sps_\perp \sum_\sv \nu_{\sv\su}^{\sps}
    s^\sv \sin(\Delta\theta_{\su\sv})\\
    & + \sin(\Delta\theta_{\sps\su}) \sum_\sv \Tilde{\psi}_{\su\sv}^{\sps} s^\sv \hat{\bm u}^\sv
    + \hat{\bm u}^\su_\perp \sum_\sv \Tilde{\lambda}_{\sv\su}^{\sps} 
    s^\sv \cos(\Delta\theta_{\sps\sv}) \\
    \Bar{\bm M}_\su^\sps & \equiv 
    \hat{\bm u}^\sps_\perp \sum_\sv \Tilde{\nu}_{\sv\su}^{\sps} 
    s^\sv \cos(\Delta\theta_{\su\sv}), \\
    \Bar{\bm N}_\su^\sps & \equiv 
    - \hat{\bm u}^\sps_\perp \sum_\sv \Tilde{\nu}_{\sv\su}^{\sps}
    s^\sv \sin(\Delta\theta_{\su\sv}).
\end{align*}

The homogeneous uniform chiral solution of Eqs.~\eqref{SM_eq_hydro_stheta} is obtained by setting gradient terms to zero and using the ansatz:
$\rho^\sps(\bm r,t) = \rho_0^\sps$,
$s^\sps(\bm r,t) = s_0^\sps$,
$\theta^\sps(\bm r,t) = \Omega t + \theta_0^\sps$,
which yields the following system of nonlinear equations
\begin{align}
    0 &= F^\sa_\sa s^\sa_0 +F^\sa_\spb {\rm C}_0 s^\spb_0 - \Tilde{F}^\sa_\spb s^\spb_0 {\rm S}_0  \nonumber \\
    \label{SM_HOMSOL}
    0 &=  F^\spb_\spb s^\spb_0 + F^\spb_\sa {\rm C}_0 s^\sa_0 + \Tilde{F}^\spb_\sa s^\sa_0 {\rm S}_0 \\
    s^\sa_0 \Omega &=  \Tilde{F}^\sa_\sa s^\sa_0 + F^\sa_\spb s^\spb_0 {\rm S}_0 + \Tilde{F}^\sa_\spb s^\spb_0 {\rm C}_0 \nonumber  \\
    s^\spb_0 \Omega & = \Tilde{F}^\spb_\spb s^\spb_0 - F^\spb_\sa s^\sa_0 {\rm S}_0 + \Tilde{F}^\spb_\sa s^\sa_0 {\rm C}_0, \nonumber
\end{align} 
where $C_0 = \cos(\theta^\spb_0 - \theta^\sa_0)$
and $S_0 = \sin(\theta^\spb_0 - \theta^\sa_0)$.
We solve these equations numerically for $s^\sa_0$, $s^\spb_0$, $\theta_0^\spb$ and $\Omega$ by fixing  
$\theta_0^\sa = 0$ without loss of generality.
In particular, we have checked that Eqs.~\eqref{SM_eq_hydro_stheta} always admit degenerate solutions symmetric under
the transformation
$(s^\sa_0,s^\spb_0,\theta_0^\spb,\Omega)
\leftrightarrow
(s^\sa_0,s^\spb_0,-\theta_0^\spb,-\Omega)
$. 

As mentioned in the main text, the homogeneous part of Eqs.~\eqref{SM_eq_hydro_stheta} also admit modulated chiral solutions for which $s^\sa$, $s\spb$, and $\theta_0^\spb - \theta_0^\sa$ exhibit weak oscillations over time. 
These solutions are generally found when the uniform chiral solution is unstable to homogeneous perturbations.
For small nonreciprocity and sufficiently large noises, we also find that that the system~\eqref{SM_HOMSOL} admits polar solutions characterized by $\Omega = \theta_0^\spb - \theta_0^\sa = 0$.
Finally, for both low noise and $\Delta\chi$, 
i.e. far away from the ordering threshold, the assumptions under which Eqs.~\eqref{SM_eq_hydro} are derived are not verified,
such that we do not find physical solutions of Eq.~\eqref{SM_HOMSOL}. 

\subsection{Floquet analysis}

We determined the linear stability of the uniform chiral solutions by performing a Floquet analysis~\cite{santoro2019introduction}. 
We start by linearizing the system around the homogeneous uniform chiral solution, which in Fourier space reads 
\begin{equation} \label{SM_linearized_eom}
    \partial_t \bm X = \mathbb{L}(t, \bm q) \bm X
\end{equation}
with $\bm X = (\delta\rho^\sa_{\bm q}, \delta\rho^\spb_{\bm q}, \delta s^\sa_{\bm q}, \delta s^\spb_{\bm q},  \delta \theta^\sa_{\bm q}, \delta \theta^\spb_{\bm q})^T$ and $\mathbb{L}(t, \bm q)$ is a linear operator periodic in time with period $T = 2\pi / \Omega$, which depends on the wavevector $\bm q$.
The formal solution of the linear system~\eqref{SM_linearized_eom} is given by
\begin{equation*}
    \bm X(t, \bm q) = \mathbb{U}(t,\bm q)\bm X(0, \bm q) ,
\end{equation*}
where $\mathbb{U}(t,\bm q)$ denotes the evolution (or Dyson) operator, which we obtain via the Dyson formula:
\begin{equation} \label{SM_opU}
    \mathbb{U}(t,\bm q) = \mathcal{T}\left\{e^{\int_0^t d\tau \mathbb{L}(\tau,\bm q)}\right\}
\end{equation} where $\mathcal{T}$ is the time-ordering operator. 
Given an initial perturbation $\bm X(0, \bm q)$, 
the uniform chiral solution is stable if and only if
the magnitude of that perturbation after one period, $|\bm X(T, \bm q)|$, is smaller than $|\bm X(0, \bm q)|$.
This amounts to require that all eigenvalues of the operator $\mathbb{U}(t,\bm q)$, which we denote as $\mathfrak{u}_i(\bm q)$, are constrained within the unit disk: $|\mathfrak{u}_i(\bm q)| < 1$.
Defining 
$$
\mathfrak{L}_i(\bm q) = \frac{1}{T} \ln \mathfrak{u}_i(\bm q),
$$
the $\mathfrak{L}_i(\bm q)$ play an analogous role as the eigenvalues of a time-independent linear evolution operatore, such that the condition for stability becomes
$$
{\rm Re}[\mathfrak{L}_i(\bm q)] \le 0, \quad \forall \bm q.
$$

In practice, we numerically evaluated the operator $\mathbb{U}(T,\bm q)$ from its definition~\eqref{SM_opU} by uniformly discretizing the time interval $[0;T]$ into $n_T$ intervals of length $\delta t = T / n_T$.
This yields the approximate formula
\begin{equation}\label{prop}
    \mathbb{U}(T,\bm q) = e^{\mathbb{L}(n_T \delta t,\bm q)\delta t} \dots e^{\mathbb{L}(2\delta t,\bm q)\delta t} e^{\mathbb{L}(\delta t,\bm q)\delta t}.
\end{equation}
We evaluated the matrix exponentials by using the\texttt{MatrixExp} function from \texttt{Mathematica}, and we typically used $n_T = 200$. Convergence was explicitly checked for multiple parameter values, and $n_T$ was found to be sufficiently large to ensure stable results. 

\subsection{Numerical integration}

Equations~\eqref{SM_eq_hydro} were numerically integrated by means of a pseudo-spectral code with explicit Euler time-stepping. 
Since the continuum model is nonlinear, we applied dealiasing using the $1/3$ rule.
All simulations are performed with spatial and temporal resolutions $dx=\tfrac{1}{2}$ and $dt=10^{-2}$, respectively.

\section{Description of the supplemental movies}

\begin{itemize}
    \item \textbf{SMov~1:} Merging of multiple bands in the solitonic band regime, indicating that one single band is the stable configuration in this region. The simulation is initialized from an existing configuration for system size $L=512$, which is duplicated 4 times to construct an initial condition containing two bands. Parameters are $\eta = 0.4$, $\Delta\chi = 0.7$ and $L=1024$. 
    \item \textbf{SMov~2:} Persistent chiral motion of particles observed when both noise and and noreciprocity are low in a small system.  Simulation parameters are $\Delta\chi = 0.2$, $\eta = 0.2$ and $L=128$. 
    \item \textbf{SMov~3:} Movie showing the nucleation of vortex-like defects, and the subsequent growth of dilute disordered bubbles. Two simulations, corresponding to two different values of nonreciprocity ($\Delta\chi = 0.1$ and $\Delta\chi = 0.3$), and therefore also two different correlation lengths are shown, illustrating that the size of these bubbles grows with the correlation length. Additional simulation parameters are $\eta = 0.2$ and $L = 2048$. 
    \item \textbf{SMov~4:} Simulation of the PDE's \eqref{SM_eq_hydro} in the metastable chiral region. Starting from a disordered initial condition, the system evolves towards a defect-choatic state. The left panel shows the relative species density $\rho^\sa - \rho^\spb$, while the left panel shows the orientation of the local momentum of $\sa$ particles. Simulation parameters are $\rho =1$, $\eta = 0.45$, $\Delta\chi = 0.35$ and $L=256$.  
\end{itemize}

Supplemental movies are available at: \url{https://owncloud.gwdg.de/index.php/s/Cit8ZLaGYBZ0aeU}.

\end{document}